%% file: iclr2025_conf.tex
\documentclass{article} 
\usepackage{iclr2025_conf,times}

\input{math_commands.tex}

\usepackage{hyperref}
\usepackage{url}
\usepackage{booktabs, multirow} 
\usepackage{graphicx}
\usepackage{subcaption}
\usepackage{amsmath,amssymb,amsfonts}

\title{Deep Reinforcement Learning for Investor-Specific Portfolio Optimization: A Volatility-Guided Asset Selection Approach}

\author{
    Arishi Orra$^{1,*}$, Aryan Bhambu$^{2}$, Himanshu Choudhary$^{1}$, Manoj Thakur$^{1}$, Selvaraju Natarajan$^{2}$ \\
    $^{1}$School of Mathematical and Statistical Sciences, Indian Institute of Technology Mandi, Mandi, India \\
    $^{2}$Department of Mathematics, Indian Institute of Technology Guwahati, Guwahati, India \\
    \texttt{arishi.orra98@gmail.com, a.bhambu@iitg.ac.in,} \\ \texttt{ch.himanshu1199@gmail.com, manoj@iitmandi.ac.in, nselvaraju@iitg.ac.in} 
}

\iclrfinalcopy 
\begin{document}

\maketitle

\begin{abstract}
Portfolio optimization requires dynamic allocation of funds by balancing the risk and return tradeoff under dynamic market conditions. With the recent advancements in AI, Deep Reinforcement Learning (DRL) has gained prominence in providing adaptive and scalable strategies for portfolio optimization. However, the success of these strategies depends not only on their ability to adapt to market dynamics but also on the careful pre-selection of assets that influence overall portfolio performance. Incorporating the investor's preference in pre-selecting assets for a portfolio is essential in refining their investment strategies. This study proposes a volatility-guided DRL-based portfolio optimization framework that dynamically constructs portfolios based on investors' risk profiles. The Generalized Autoregressive Conditional Heteroscedasticity (GARCH) model is utilized for volatility forecasting of stocks and categorizes them based on their volatility as aggressive, moderate, and conservative. The DRL agent is then employed to learn an optimal investment policy by interacting with the historical market data. The efficacy of the proposed methodology is established using stocks from the Dow $30$ index. The proposed investor-specific DRL-based portfolios outperformed the baseline strategies by generating consistent risk-adjusted returns.
\end{abstract}

\section{Introduction} \label{Introduction}

\input{Introduction}

\section{Problem Formulation} \label{Problem}
\input{Problem}

\section{Proposed Methodology} \label{Methodology}
\input{Methodology}

\section{Experiments} \label{Results}
\input{Results}

\section{Conclusion} \label{Conclusion}
\input{Conclusion}


\input{iclr2025_conf.bbl}
\end{document}

%% file: math_commands.tex
\usepackage{amsmath,amsfonts,bm}

\def\eqref#1{equation~\ref{#1}}

\def\1{\bm{1}}

\DeclareMathAlphabet{\mathsfit}{\encodingdefault}{\sfdefault}{m}{sl}
\SetMathAlphabet{\mathsfit}{bold}{\encodingdefault}{\sfdefault}{bx}{n}



%% file: Introduction.tex
Recent studies have shown the potential of Deep Reinforcement Learning (DRL) in dynamic portfolio allocation. DRL enables sequential decision-making through trial-and-error interactions with the stock market environment \citep{sutton2018reinforcement}. DRL uses deep learning to detect complex market patterns, offering a flexible and scalable solution for portfolio optimization problems. \cite{moody1998performance} reported the first instance of employing RL utilizing recurrent RL for portfolio optimization. After that, numerous research has been conducted on the use of DRL in portfolio optimization \citep{jiang2017deep, aboussalah2020continuous, wang2021deeptrader, jang2023deep, cui2024multi}.


The choice of an optimal set of assets is crucial for constructing a portfolio as it determines its overall performance. Effective pre-selection helps to reduce computational efficiency, avoids the risk of overfitting, and ensures proper diversification. However, very few studies focus on the crucial component of asset pre-selection before forming the portfolio. Aligning the asset pre-selection with investor preferences is essential in refining investment strategies \citep{wang2020portfolio}. These refined strategies meet individual risk tolerances and return expectations. Generally, three investor types exist in the market: Aggressive, Moderate, and Conservative \citep{yadav2024multiobjective}. Selecting assets that align with an investor's risk profile enables the construction of portfolios that optimize returns while maintaining adherence to predefined risk parameters. This increases the likelihood of achieving the individual's investment goals. 

This study proposes a volatility-guided DRL-based portfolio optimization framework that dynamically constructs portfolios based on investors' risk profiles. The proposed approach leverages the Generalized Autoregressive Conditional Heteroscedasticity (GARCH) \citep{satchell2011forecasting} model for stock volatility forecasting and classifies them according to their volatility. Based on an investor's risk appetite, we construct three distinct portfolios containing Aggressive, Moderate, and Conservative stocks to fulfill their individual investment goals. A DRL-based agent is then employed to learn an optimal investment policy by interacting with the historical market data. The efficacy of the proposed methodology is established using stocks from the Dow $30$ index. The proposed three investor-specific DRL-based portfolios outperformed the baseline strategies by generating consistent risk-adjusted returns.

The remainder of the paper is organized as follows. Section \ref{Methodology} formulates the portfolio optimization problem into the DRL setup and section \ref{Problem} introduces the proposed methodology. Section \ref{Results} covers the data description and experimental settings, and discusses the findings of the experiment. Finally, section \ref{Conclusion} draws meaningful conclusions from the study.

%% file: Problem.tex
    Portfolio optimization refers to the dynamic allocation of capital into a variety of assets with the goal of achieving long-term cumulative returns with minimal risk. The dynamic and stochastic nature of the financial markets offers the flexibility to model it as a Markov Decision Process (MDP). The portfolio optimization problem in an MDP formulation is defined as the tuple $(S, A, \mathbb{P}, R, \gamma)$, where $S$ and $A$ are the collections of states and actions respectively, $\mathbb{P}$ represents the state transition probability, $R$ is the reward distribution and $\gamma \in [0,1]$.

    At each time step $t$, state $s_t$ comprises the prices of the stocks, the covariance matrix of the prices of the stocks, and some commonly used technical indicators. Action $a_t$ is a $n$-dimensional vector representing the amount of funds to be allocated to each of the $n$-assets, and reward $r_t$ is the portfolio return attained after executing $a_t$ in the environment. This notion of states and environment has been adopted from \citep{liu2020finrl}. The objective is to find an optimal policy, $\pi$, determining the optimal allocation of funds by interacting with the environment.

%% file: Methodology.tex
    Asset selection for a portfolio is crucial as it balances the risk-return tradeoff and directly influences the portfolio's performance. Determining an ideal set of assets presents several challenges, such as diversification, market volatility, liquidity constraints, etc. This study proposes a volatility-guided stock pre-selection framework for portfolio optimization using DRL. The proposed approach dynamically adjusts asset allocation based on market conditions and constructs portfolios according to investors' risk profiles. A pool of stocks is selected, and the GARCH model is applied to estimate the volatility of each stock to provide a data-driven foundation for portfolio construction.

    The GARCH \citep{bollerslev1986generalized, satchell2011forecasting} model has been utilized to verify the volatility of the stocks in the index due to its effectiveness in capturing time-dependent volatility patterns. The GARCH$(1,1)$ model is represented as:
    \begin{equation}
    r_t = \mu_t + \epsilon_t, \quad \epsilon_t \sim N(0, \sigma_t^2)
    \label{eqn:g1}
    \end{equation} 
    Here, $r_t = \log \left(\frac{p_t}{p_{t-1}}\right)$ represents the logarithmic returns at time $t$ and $p_t$ is the price at time $t$.  The mean equation of the return in Equation (\ref{eqn:g1}) is assumed to be zero. Then, the GARCH$(1,1)$ model is mathematically expressed as:
    \begin{equation}
       \sigma_t^2 = \omega + \alpha \epsilon_{t-1}^2 + \beta \sigma_{t-1}^2
    \end{equation} 
    Here, $\epsilon_{t-1}^2$ represents past squared shocks and $\sigma_{t-1}^2$ represents and influences the future volatility $\sigma_t^2$. Their impact decreases with time due to the autoregressive structure, leading to a short memory effect, with the stationarity ensured by $0 < \alpha + \beta < 1$. This formulation highlights how past returns and variances influence current volatility, with autoregressive and moving average components governing its persistence.


    Based on the volatility estimates from the GARCH model, the pool of stocks is classified into three distinct sets of assets:
    \begin{itemize}
        \item \textbf{Aggressive}: Containing the most volatile stocks, suitable for investors willing to take high-risk chasing potentially higher returns.
        \item \textbf{Moderate}: Consisting of stocks with medium volatility, providing a balance between risk-return tradeoffs.
        \item \textbf{Conservative}: Comprising of the least volatile stocks, tailored for risk-averse investors looking for more stable returns.
    \end{itemize}

\begin{figure}[!htbp]
        \centering
        \includegraphics[scale=0.35]{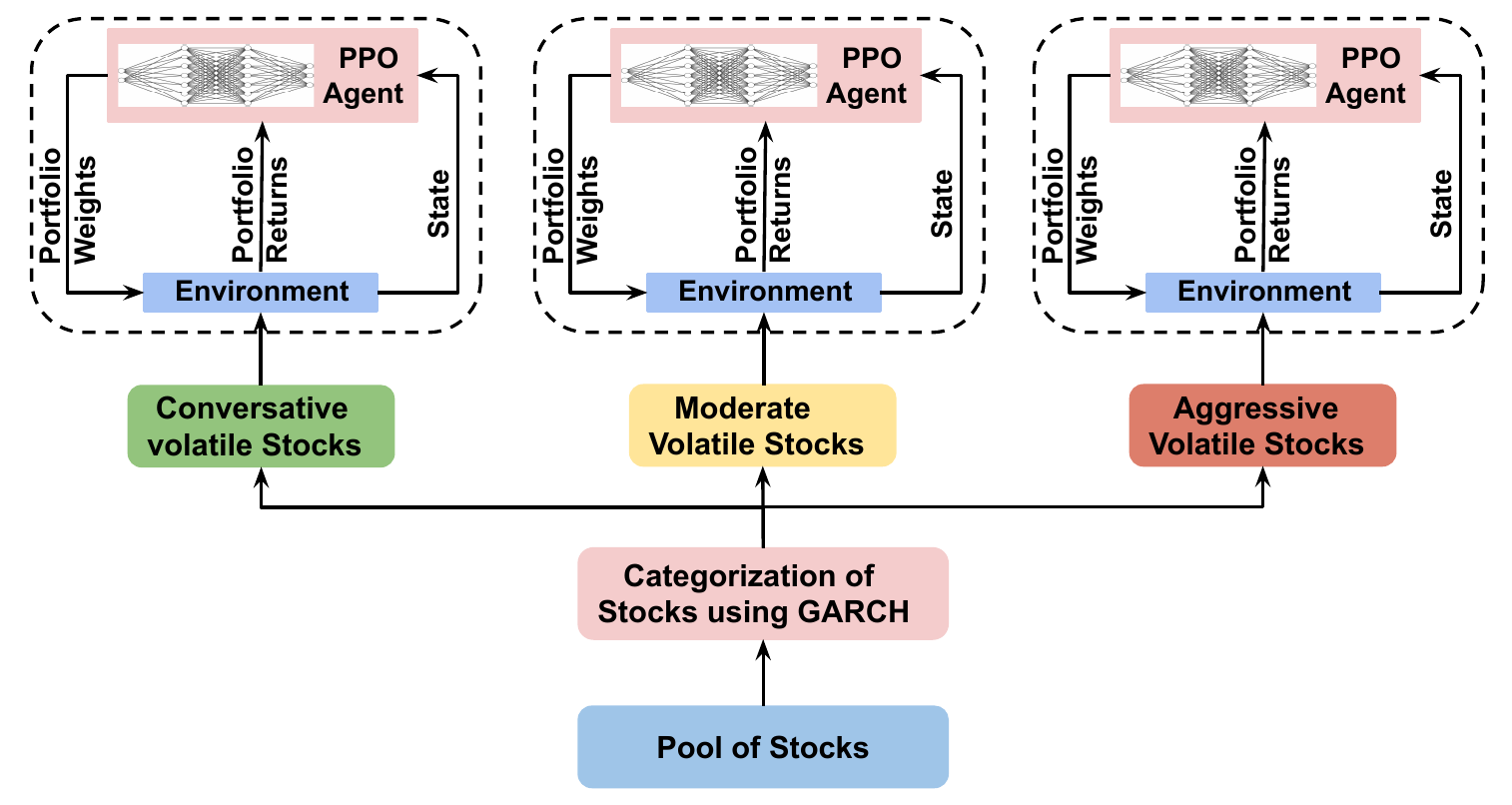}
        \caption{The schematic diagram depicting the workflow of the proposed methodology.}
        \label{fig:methodology}
    \end{figure}

    According to their risk-taking capability, the investors can choose their class of assets and invest in them to fulfill their investment goals. Finally, a DRL-based agent is utilized to optimize the weights of these portfolios. The complete workflow of the proposed approach is depicted in Figure (\ref{fig:methodology}). In particular, at each step, GARCH classifies the stocks into three distinct sets, and a Proximal Policy Optimization (PPO) \citep{schulman2017proximal} agent is employed to allocate the portfolio weights optimally.

%% file: Results.tex
\subsection{Data Description and Experimental settings}

    The proposed methodology has been evaluated using the data from the Dow Jones Index (U.S.). Daily price data for all $30$ stocks of the DJI index was obtained from \href{https://finance.yahoo.com}{Yahoo Finance} over a period from January $1$, $2010$, to December $31$, $2024$. The training data span from January $1$, $2010$, to December $31$, $2022$, and the data from January $1$, $2023$, to December $31$, $2024$ is utilized for out-of-sample evaluation. The top $10$ volatile stocks of the DJI index are used to build an aggressive portfolio, the least volatile $10$ are used to build a conservative portfolio, and the remaining are used to develop a moderate portfolio.

    To simulate real-world trading conditions, an initial capital of \$1 million has been provided to the agent for investing in selected stocks. Motivated from \cite{orra2024dynamic}, a transaction cost of $0.05\%$ has been applied to all trades to reflect market conditions accurately. Bayesian optimization has been utilized to fine-tune the hyperparameters of the DRL models, following the methodology outlined in \citep{bhambu2024recurrent} to ensure that the models were well-calibrated, thereby enhancing their decision-making capabilities in portfolio management. The search ranges for hyperparameter tuning were determined empirically, facilitating the efficient exploration of the parameter space while minimizing computational overhead.

\subsection{Results and Discussion}


    The effectiveness of the proposed volatility-guided investor-specific portfolios is evaluated by comparing its performance against three benchmark strategies: The Mean-Variance Portfolio Optimization (MVO) strategy \citep{markowitz1991foundations}, the DJI index, and the Equal Weighted Portfolio. To ensure a comprehensive evaluation, we have mainly selected five risk and return performance metrics to assess the results: Annual Returns, Cumulative Returns, Sharpe Ratio, Maximum Drawdown, and Annual Volatility. Table (\ref{tab:table1}) details the performance comparison of the proposed approach versus the benchmark strategies. All DRL-based portfolio optimization models were trained five times, and their average results were recorded.

    \begin{table}[!htp]\centering
    \caption{Performance evaluation of the proposed methodology against the benchmarks portfolio strategies.}
    \label{tab:table1}
    \resizebox{\textwidth}{!}{
    \begin{tabular}{lcccccc}\toprule
    \textbf{Models} &\textbf{Annual Return} &\textbf{Cumulative Return} &\textbf{Sharpe Ratio} &\textbf{Max Drawdown} &\textbf{Annual Volatility} \\\midrule
    Aggressive-DRL &\textbf{26.714} &\textbf{60.629} &1.216 &15.596 &16.358 \\
    Moderate-DRL &24.002 &53.372 &\textbf{1.811} &10.335 &12.323 \\
    Conservative-DRL &19.461 &42.816 &1.559 &9.969 &\textbf{10.078} \\
    MVO \citep{markowitz1991foundations} &7.718 &15.896 &0.828 &9.536 &10.517 \\
    DJI &14.023 &29.743 &1.211 &9.017 &11.387 \\
    Equal-Weighted &13.462 &28.479 &1.178 &\textbf{8.676} &11.264 \\
    \bottomrule
    \end{tabular} }
    \end{table}

    The Aggressive-DRL portfolio, consisting of highly volatile stocks, achieved the highest annual ($26.71\%$) and cumulative ($60.63\%$) returns among all the models. However, these returns come at the cost of increased risk regarding high annual volatility and drawdown. On the other hand, the Conservative-DRL achieves the least volatility ($10.08\%$) and a lower drawdown ($9.97\%$), showing its effectiveness in optimizing portfolio risk. The Moderate-DRL portfolio balances the risk and return and delivers significant returns with lower risk measures. Also, it attains the highest Sharpe ratio ($1.81$), indicating a better risk-adjusted performance among all the other models. All the DRL-based portfolio models significantly outperformed the benchmark models, which considered the entire pool of stocks for constructing the portfolios. This highlights the efficacy of our proposed methodology in pre-selecting assets according to investors' risk preferences and delivering consistent performance using DRL.

    \begin{figure}[!htbp]
        \centering
        \includegraphics[scale=0.18]{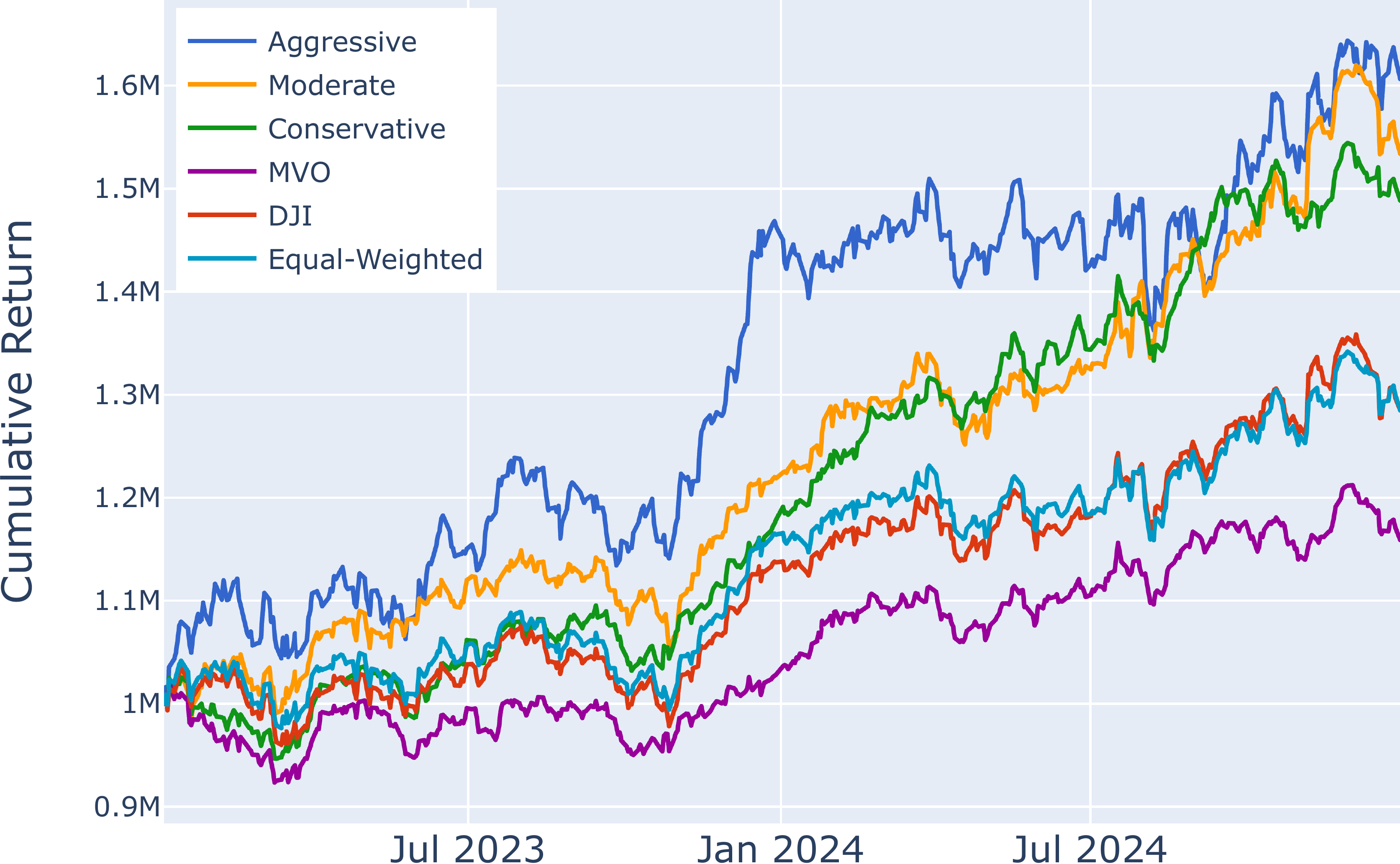}
        \caption{Cumulative return plots of the proposed methodology against the benchmark portfolio strategies over the trading period.}
        \label{fig:cum_plots}
    \end{figure}

    Figure (\ref{fig:cum_plots}) illustrates the cumulative return plot of the three investor-specific DRL-based portfolios and the benchmark strategies. The Aggressive-DRL portfolio exhibits the highest growth but with larger fluctuations, indicating its vulnerability to volatile stocks. The Moderate-DRL portfolio shows strong performance with fewer setbacks, demonstrating a balanced performance. Although the Conservative-DRL produced marginally lower returns, it delivered a stable performance. Overall, the Moderate-DRL portfolio emerges as the most balanced strategy and is ideal for investors looking to balance the risk-return tradeoff of the investment. On the other hand, the Aggressive and Conservative portfolios are well suited for risk-seeking and risk-averse investors, respectively.

%% file: Conclusion.tex

This study proposes a novel methodology by harnessing the strengths of GARCH and DRL to construct volatility-guided investor-specific portfolios. The proposed framework categorizes the stocks based on their volatility and dynamically constructs three distinct portfolios according to investors' risk appetite. The proposed methodology is assessed using several risk and return performance measures for the Dow Index. The Aggressive-DRL portfolio attained the highest returns with increased volatility, which is suitable for risk-seeking investors. The Moderate-DRL portfolio achieved the most balanced performance, while Conservative-DRL provided stable returns with minimal volatility. The proposed volatility-guided DRL-based portfolios significantly outperformed the benchmark strategies considering various risk-return measures. Future research could explore dynamic risk measures such as CVaR, expanding to multi-asset class portfolios, and including macroeconomic indicators to improve the model's robustness.